\documentclass[twocolumn,aps,showpacs,pre]{revtex4}

\usepackage{graphicx}
\usepackage{epsfig}
\usepackage{dcolumn}
\usepackage{amsfonts}
\usepackage{amsmath}
\usepackage{amssymb}
\usepackage{bm}
\usepackage{longtable}

\begin{document}

\newcommand{\tens}[1]{\tensor{#1}}
\newcommand{\brm}[1]{\bm{{\rm #1}}}

\title{Multifractality in a broad class of disordered systems}

\author{Olaf Stenull}

\affiliation{
Department of Physics and Astronomy, University of Pennsylvania, Philadelphia, PA 19104}
\date{\today}

\begin{abstract}
We study multifractality in a broad class of disordered systems which includes, e.g., the diluted $x$-$y$ model. Using renormalized field theory we analyze the scaling behavior of cumulant averaged dynamical variables (in case of the $x$-$y$ model the angles specifying the directions of the spins) at the percolation threshold. Each of the cumulants has its own independent critical exponent, i.e., there are infinitely many critical exponents involved in the problem. Working out the connection to the random resistor network, we determine these multifractal exponents to two-loop order. Depending on the specifics of the Hamiltonian of each individual model, the amplitudes of the higher cumulants can vanish and in this case, effectively, only some of the multifractal exponents are required. 
\end{abstract}

\pacs{64.60.Ak}

\maketitle

The physics of critical phenomena is usually characterized by a few critical exponents. In certain, e.g. nonlinear, systems, however, the scaling behavior can be much richer. It can be even so complex, that its exhaustive characterization requires infinitely many critical exponents~\cite{halsey&Co_86}. This is the famous phenomenon of multifractality~\cite{review_multifractality}. Recently studied examples are as diverse as heartbeat~\cite{ivanov_99}, quantum gravity~\cite{duplantier_99} and percolation~\cite{bunde_havlin_91_stauffer_aharony_92} type problems like random resistor networks (RRNs)~\cite{stenull_janssen_rrn_multifract} and self-avoiding walks on percolation clusters~\cite{vonFerber&Co_2003}.

In this note we study the multifractal properties of a broad class of diluted physical systems, viz.\ those systems that can be described by lattice models with a Hamiltonian that is a sum of bond energies $U$ depending only on the differences $\vartheta_{i,j} = \vartheta_i - \vartheta_j$ of continuous dynamical variables $\vartheta_i$ and $\vartheta_j$ on the bonds $\langle i,j\rangle$ between nearest neighboring sites $i$ and $j$,
\begin{subequations} 
\label{typeOfH}  
\begin{eqnarray}
H = \sum_{\langle i,j\rangle} \gamma_{i,j} \, U \left( \vartheta_{i,j}  \right) \, .
\end{eqnarray}
Here, $\gamma_{i,j}$ is a random variable that mimics disorder. It is assumed to take on the values $1$ and $0$ with respective probabilities $p$ and $1-p$. We focus on systems that are macroscopically isotropic and hence the bonds are assumed to be undirected, $U ( \vartheta ) = U ( -\vartheta )$. Moreover, $U$ is assumed to have a well defined minimum about which it can be expanded in a power series in $\vartheta$ (the locus of this minimum is used to define $\vartheta=0$). Otherwise, $U$ is arbitrary. For example, it may be periodic or not.  Given these assumptions and conventions, the bond energy has a Taylor expansion of the form
\begin{eqnarray}
\label{bondTaylor}
U(\vartheta)  = \sum_{l=0}^\infty a_l \, \vartheta^{2l} 
\end{eqnarray}
\end{subequations}
with $a_1$ being strictly positive. Diverse physical systems can be described by this type of Hamiltonian. The simplest of these systems is perhaps the RRN, where $\vartheta_i$ corresponds to the voltage $V_i$ at site $i$ and is defined on the interval $[-\infty, \infty]$. $U(V) = \frac{1}{2} \sigma V^2$, with $\sigma$ being the bond conductance, is the electric power dissipated on an occupied bond. A whole family of systems that can be described by the Hamiltonian $H$ is the family of systems with $x$-$y$ symmetry, i.e., systems that are invariant under the orthogonal group $O_2$ of rotations in a two-dimensional plane and the isomorphic group $U(1)$ of transformations of the phase of a complex number. The most intuitive example here is perhaps a ferromagnet in which the spins are confined by crystal fields to lay in a certain plane. Other systems exhibiting $x$-$y$ symmetry include superconductors, superfluid helium, the smectic-C and the hexatic-B phase of liquid crystals and so on. In the $x$-$y$ model $\vartheta_i$ becomes the angle $\varphi_i$ that specifies the orientation of the spin at site $i$ and is defined on the interval $[-\pi , \pi]$. The bond energy is a $2\pi$-periodic function, $U(\varphi) = - K \cos ( \varphi)$, with $K$ being the exchange integral.

Since we are dealing with diluted systems, we are facing basically a percolation problem. If $p$ is small, there are only finite clusters. If $p$ exceeds a certain threshold value $p_c$, on the other hand, there exists an infinite cluster. At the threshold, $p=p_c$, the system undergoes an isotropic percolation (IP) transition. Hence, the order parameter $P_\infty$ (the probability that any site belongs to an infinite cluster) and the correlation length $\xi$ (the average diameter of a finite cluster) scale as $P_\infty \sim (p-p_c)^\beta$ and $\xi \sim |p-p_c|^{-\nu}$, respectively, where $\beta$ and $\nu$ are the well known critical exponents of the IP universality class. Here, we are interested primarily in  physical processes taking place on the clusters like electric conduction or the interaction of spins. We will see that the cumulants
\begin{equation}
\label{defCphi}
C_\vartheta^{(l)}  (x, x^\prime) = \big\{ \left\langle   \vartheta_{x, x^\prime} ^{2l} \right\rangle_c  \big\}_{\mathfrak{C}}^\star 
\end{equation}
are adequate and convenient observables to investigate the multifractality of such processes.  $\langle \cdots \rangle_c$ stands for the cumulants of the average $\langle \cdots \rangle$ with respect to the Hamiltonian~(\ref{typeOfH}), e.~g., $\langle \vartheta^2 \rangle_c = \langle \vartheta^2 \rangle$, $\langle \vartheta^4 \rangle_c = \langle \vartheta^4 \rangle - 3 \langle \vartheta^2 \rangle^2$ and so on. $\{ \cdots \}_{\mathfrak{C}}$ denotes averaging over all configurations of the diluted lattice and the star indicates the constraint that $x$ and $x^\prime$ must be connected.

To our knowledge, the scaling behavior of the cumulants~(\ref{defCphi}) is not known to date with 2 exceptions: (i) the RRN where one has conventional gap scaling because $H$ is harmonic, $C_\vartheta^{(l)}  (x, x^\prime) \sim |x - x^\prime|^{l \phi/\nu}$, with $\phi$ being the resistance exponent known to second order in $\varepsilon$ expansion and (ii) the diluted $x$-$y$ model where $C_\vartheta^{(1)}  (x, x^\prime) \sim |x - x^\prime|^{\phi/\nu}$ and $C_\vartheta^{(2)}  (x, x^\prime) \sim |x - x^\prime|^{\phi_c/\nu}$ with $\phi_c$ being a critical exponent associated with corrections to scaling that is known to first order in $\varepsilon$ expansion~\cite{harris_lubensky_87b,janssen_stenull_irrelevant_2004}. The purpose of this note is to reveal the scaling behavior of the $C_\vartheta^{(l)}$ {\em for  all systems} covered by the Hamiltonian~(\ref{typeOfH}) {\em for  all $l$}. Using renormalized field theory we will explore the intricate connection of the present problem to the renowned noisy RRN~\cite{noisyRRN}  {\em to arbitrary order} in perturbation theory. We will show that the $C_\vartheta^{(l)} $ scale at the percolation threshold as
\begin{equation}
\label{scaleC}
C_\vartheta^{(l)}  (x, x^\prime) = A_l \,  |x - x^\prime|^{\psi_l/\nu}  ,
\end{equation}
with the exponents $\psi_l$ being identical to the noise exponents of the RRN~\cite{park_harris_lubensky_87} and with the $A_l$ being amplitudes which depend on the specifics of $U$, i.e., $A_l \sim a_l$.

{\em Field theoretic model.}--In order to apply field theory and renormalization group (RG) methods we need to condense the Hamiltonian~(\ref{typeOfH}) into a field theoretic Hamiltonian that is suitable for studying the $C_\vartheta^{(l)}$. This can be done by following the seminal work of Harris and Lubensky (HL) on the RRN and the diluted $x$-$y$ model~\cite{harris_lubensky_87b} with the result 
\begin{eqnarray}
\label{HLH}
\mathcal{H} = \int d^d x \int_{\vec{\theta}}\bigg\{ \frac{1}{2}\,  \Phi K \left( \nabla, \nabla_{\vec{\theta}} \right) \Phi + \frac{g}{6} \, \Phi^3\bigg\} \, ,
\end{eqnarray}
where the Gaussian kernel is given by
\begin{eqnarray}
\label{kernel}
K \left( \nabla, \nabla_{\vec{\theta}} \right) = \tau - \Delta - w \Delta_{\vec{\theta}} - \sum_{\l=2}^\infty v_l \sum_{\alpha =1}^D \left( \nabla_{\theta^{(\alpha )}}^2 \right)^l \, .
\end{eqnarray}
The order parameter field $\Phi({\bf x},\vec{\theta})$ lives on a continuous $d$-dimensional space with the coordinates ${\bf x}$. It is subject to the constraint $\int_{\vec{\theta}}\Phi({\bf x},\vec{\theta})=0$. The variable $\vec{\theta}$ is a replicated analog of the dynamic variable $\vartheta$ and lives on a $D$-dimensional torus~\cite{fotenote_regularization}. The physical situation is recovered in the replica limit $D\to 0$. The parameter $\tau$ is proportional to $p_c -p$, i.e., it specifies the distance from the critical point. $w$ is proportional to $a_1 $ and $v_l \sim a_l$. For vanishing $v_l$ the Hamiltonian~(\ref{HLH}) reduces to the original field theoretic Hamiltonian of HL. The $v_l$ are dangerous irrelevant couplings as far as the $C_\vartheta^{(l)}$ are concerned. This can be seen by performing a scaling analysis in the replica variable $\vec{\theta}$ that leads to
\begin{equation}
\label{formC}
C_\vartheta^{(l)}  (x, x^\prime; \tau , w , \left\{ v_k \right\}) = w^l f_l  (x, x^\prime; \tau , \left\{ v_k /w^k \right\}) \, ,
\end{equation}
where $f_l$ is a scaling function. This shows that the $v_l$ exclusively appear in the irrelevant combination $v_l /w^l$. However, it turns out the leading contribution to $C_\vartheta^{(l)}$ vanishes upon setting $v_l $ to zero, i.e., information about the leading scaling behavior of $C_\vartheta^{(l)}$ is lost by omitting $v_l$ and this is why the $v_l$ are dangerous.

{\em Physical contents.}--To fully appreciate the physical contents of the Hamiltonian~(\ref{HLH}) it is helpful to consider the replica space Fourier transform $\psi_{\vec{\lambda}} (\brm{x}) =  \int_{\vec{\theta}} \exp ( - i\vec{\lambda} \cdot \vec{\theta})  \Phi({\bf x},\vec{\theta})$ of the order parameter, where $\vec{\lambda}$ is the replica variable conjugate to $\vec{\theta}$. $\vec{\lambda}$ takes on values on a discrete $D$-dimensional torus. The quantity $\psi_{\vec{\lambda}} (\brm{x})$ is designed so that its correlation function
\begin{equation}
\label{corrPsi}
G \left( \brm{x}, \brm{x}^\prime ; \vec{\lambda} \right) = \left\langle  \psi_{\vec{\lambda}} (\brm{x}) \psi_{- \vec{\lambda}} (\brm{x}^\prime )\right\rangle_{\mathcal{H}} ,
\end{equation}
where $\langle \cdots \rangle_{\mathcal{H}}$ indicates averaging with respect to the Hamiltonian~(\ref{HLH}), provides convenient access to the cumulants $C_\vartheta^{(l)}$. Applying a standard cumulant expansion one finds
\begin{align}
\label{corrPsiXY}
G \left( \brm{x}, \brm{x}^\prime ; \vec{\lambda} \right) = 
  \left\{ \exp \left[ \sum_{l=1}^\infty \frac{(-1)^l}{(2l)!} \, K_l(\vec{\lambda}) \left\langle \vartheta_{\brm{x},\brm{x}^\prime}^{2l} \right\rangle_c \right] \right\}_{\mathfrak{C}}  
\end{align}
where $K_l (\vec{\lambda}) = \sum_{\alpha =1}^D [\lambda^{(\alpha)}]^{2l}$ is homogeneous polynomial in $\vec{\lambda}$ of degree $2l$. Equation~(\ref{corrPsiXY}) shows that $C_\vartheta^{(l)}$ can be calculated via taking the derivative with respect to $K_l (\vec{\lambda})$, or in other words, that $G ( \brm{x}, \brm{x}^\prime ; \vec{\lambda} )$ is a generating function for the $C_\vartheta^{(l)}$. This property will play an important role as we go along; it will alow us to extract the scaling behavior of the $C_\vartheta^{(l)}$ from that of $G ( \brm{x}, \brm{x}^\prime ; \vec{\lambda} )$ which in turn can be calculated by using field theory and RG methods. 

{\em Diagrammatic perturbation theory.}--As usual, the central element of our RG analysis is a diagrammatic perturbation calculation. Its constituting elements are the three-leg vertex $-g$ and the Gaussian propagator $G ( \brm{k}, \vec{\lambda} ) \{ 1 - \delta_{\vec{\lambda}, \vec{0}} \}$, where $G (\brm{k}, \vec{\lambda}) = (\tau + \brm{k}^2 + w \vec{\lambda}^2)^{-1}$ and where $\brm{k}$ is a momentum or wave vector conjugate to $\brm{x}$. Due to the factor $\{ 1 - \delta_{\vec{\lambda}, \vec{0}} \}$, which enforces the constraint $\psi_{\vec{0}} (\brm{x}) =  0$ stemming from $\int_{\vec{\theta}}\Phi({\bf x},\vec{\theta})=0$, the principal propagator decomposes in a replica carrying part $G ( \brm{k}, \vec{\lambda})$ and a part $G ( \brm{k}, \vec{\lambda})\delta_{\vec{\lambda}, \vec{0}}$ not carrying replica variables. Each principal diagram decomposes into a sum of replica carrying diagrams consisting of these two types of propagators. 

Note that none of the irrelevant $v_l$ appears in the propagators. This is important because treating the $v_l$ in the same way as the relevant couplings $\tau$ and $w$ would ruin our perturbation expansion, i.e., increasing orders in an expansion of the Feynman diagrams in terms of the $v_l$ lead to increasing superficial degrees of divergence. It is mandatory to truncate this expansion, or in other words, we should treat the $v_l$ by means of insertions of the composite field (or briefly operator)
\begin{eqnarray}
\mathcal{O}_l(\brm{x})= \frac{v^l}{2} \int_{\vec{\theta}} \Phi({\bf x},\vec{\theta})   \sum_{\alpha =1}^D \left( \nabla_{\theta^{(\alpha )}}^2 \right)^l \Phi({\bf x},\vec{\theta})   \, .
\end{eqnarray} 

For the following arguments it is useful to employ the so-called Schwinger parametrization, i.e., to rewrite the propagators by using the mathematical identity $G (\brm{k}, \vec{\lambda}) = \int_0^\infty ds \, \exp[  - s (\tau + \brm{k}^2 + w \vec{\lambda}^2) ]$. Let us consider a generic replica carrying  Feynman diagram with successive single insertions of $\mathcal{O}_l$ in each of its replica carrying propagators. The $\vec{\lambda}$-dependent part of such a diagram is of the form
\begin{eqnarray}
 v^l \sum_i s_i  \sum_{\{ \vec{\kappa} \} } K_l \left( \vec{\lambda}_i \right) \, \exp \bigg( w \sum_j s_j \vec{\lambda}_j^2\bigg)  \, .
\end{eqnarray}
Here, the summation $\sum_{\{ \vec{\kappa} \} }$ is a summation over some complete set of independent loop replica variables $\{ \vec{\kappa} \}$. The summations indexed by $i$ and $j$ are taken over all the replica carrying propagators. $\vec{\lambda}_j = \vec{\lambda}_j (\{ \vec{\kappa} \}, \vec{\lambda})$, where $\vec{\lambda}$ denotes an external replica variable, is the total replica variable flowing through propagator $j$. The summation over $\{ \vec{\kappa} \}$ can be simplified by a completion of squares in the exponential and eventually approximated by an integration. This integration is Gaussian and hence straightforward. Taking the replica limit $D\to 0$ and using that $K_l (\vec{\lambda})$ is a homogeneous polynomial of degree $2l$ one obtains
\begin{align}
\label{finalLambdaPart}
   v^l \sum_i s_i \, c_i \left( \left\{ s \right\} \right)^{2l} K_l \left( \vec{\lambda} \right) 
 + \cdots\, , 
\end{align}  
where $c_i ( \{ s \} )$ is a homogeneous function of the Schwinger parameters $ \{ s \}$ of degree zero that depends exclusively on the topology of our generic diagram.

The remaining steps of calculating our generic Feynman diagram consist of integrating out the loop momenta and the Schwinger parameters. These steps are entirely analogous to those well known from the field theory of IP and are skipped here for briefness. For background of the methods involved here, e.g, dimensional regularization and minimal subtraction involving Laurent expansions in $\varepsilon = 6-d$, we refer to Ref.~\cite{amit_zinn-justin}. 

Beyond these standard procedures there is one intricacy involved here that warrants further comment. The ellipsis in expression~(\ref{finalLambdaPart}) stands for various terms each of which contains a homogeneous polynomial in $\vec{\lambda}$. The polynomials of the omitted superficially divergent terms, however, all have a higher symmetry than $K_l(\vec{\lambda})$. If an operator which depends on $\vec{\lambda}$ via one of the more symmetric polynomials is inserted into one of the IP Feynman diagrams it can generate all sorts of polynomials, or for that matter operators, but it can never generate $\mathcal{O}_l$. This feature distinguishes $\mathcal{O}_l$ and makes $\mathcal{O}_l$ a master operator~\cite{stenull_janssen_rrn_multifract} whereas the other operators are just slaves. All slaves must be taken into account  in the renormalization process and one has, at least in principle, to deal with entire renormalization matrixes instead of simple renormalization factors. However, these renormalization matrices have a particular, simple structure. Due to this simple structure, the scaling exponent of a master operator such as $\mathcal{O}_l$ is completely determined by a single element of the renormalization matrix. Hence, for the practical purpose of calculating a masters scaling exponent, the slaves can be neglected.

{\em Comparison to the noisy random resistor network.}--The noisy RRN is a generalization of the RRN in which the conductances of occupied bonds are random variables. Thus, there are two types of quenched disorder in this model, viz. the dilution and the randomness of the conductance of individual bonds. To treat the two types of disorder simultaneously, Park, Harris and Lubensky (PHL)~\cite{park_harris_lubensky_87} introduced a variant of the HL model in which the role of the $D$-fold replicated  $\vec{\theta}$ is taken by a $(D\times E)$-fold replicated voltage $\tens{\theta}$. The perturbation theory for the PHL model can be performed essentially by following the steps described above. The only noteworthy difference is that the field theoretic operators $\mathcal{O}_l^{\text{RRN}}$ leading to the noise exponents describing the current distribution on the network contain instead of $K_l (\vec{\lambda})$ the polynomials $K_l^{\text{RRN}} (\tens{\lambda}) = \sum_{\beta =1}^E [ \sum_{\alpha =1}^D (\lambda^{(\alpha, \beta)})^2 ]^l$ , where $\tens{\lambda}$ is the replica current conjugate to the replica voltage $\tens{\theta}$. Though different, $K_l (\vec{\lambda})$ and $K_l^{\text{RRN}} (\tens{\lambda})$ share two pivotal properties. First, they are of sufficiently low symmetry so that the corresponding operators $\mathcal{O}_l$ and $\mathcal{O}_l^{\text{RRN}}$are master operators. Second, both are homogeneous polynomials of degree $2l$. Thus, in both cases the perturbation theory leads to expression~(\ref{finalLambdaPart}) [of course with either $K_l (\vec{\lambda})$ or $K_l^{\text{RRN}}(\tens{\lambda})$] up to unimportant differences residing in the ellipsis, i.e., up to different slaves. Therefore, the two perturbation theories lead to identical results as far as the scaling behavior of the master operators is concerned.

\begin{figure}
\includegraphics[width=7.5cm]{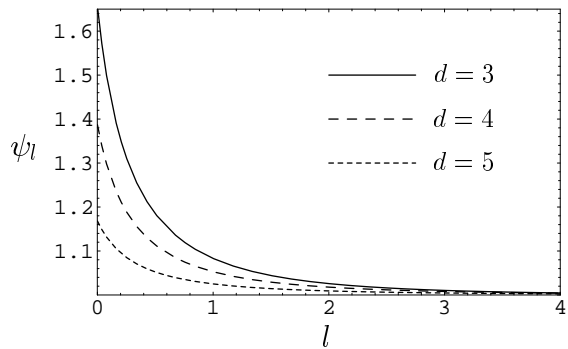} 
\caption{Dependence of the multifractal exponents $\psi_l$ on $l$ in three, four and five dimensions.}
\label{curves}
\end{figure}

{\em Scaling behavior.}--Having made this observation we can draw on the noisy RRN, in particular on Refs.~\cite{stenull_janssen_rrn_multifract}, for the remaining steps. Eventually we are led for the correlation functions at criticality to the scaling form
\begin{align}
\label{scalingForm}
G \left( {\bf x}, {\bf x}^\prime ;\vec{\lambda} \right) & = |{\bf x}-{\bf x}^\prime |^{-2\beta/\nu} \Big\{ B_0 + 
\nonumber \\
&+ \sum_{l=1}^\infty B_l v_l K_l \left( \vec{\lambda} \right) |{\bf x}-{\bf 
x}^\prime |^{\psi_l/\nu} + \cdots \Big\} \, .
\end{align}
In writing Eq.~(\ref{scalingForm}) we have used that $K_1 ( \vec{\lambda} ) = \vec{\lambda}^2$ and we have set $w = v_1$. The $B$'s are expansion coefficients. The multifractal exponents $\psi_l$  are identical to the noise exponents of the RRN and hence they are known to second order in $\varepsilon$~\cite{stenull_janssen_rrn_multifract},
\begin{widetext}
\begin{align}
\label{monsterExponent}
\psi_l &= 1 + \frac{\varepsilon}{7  \left( 1+l \right) \left( 1+2l \right)} + 
  \frac{ 313 + l\,\left\{ 3327 + 
          8\,l\,\left[ 1556 + l\,\left( 2076 + 881\,l \right)  \right]  \right\}  - 
       672\,{\left( 1 + l \right) }^2\,{\left( 1 + 2\,l \right) }^2\,
        H(2\,l)  }{12348\,{\left( 1 + l \right) }^3\,
     {\left( 1 + 2\,l \right) }^3} \, \varepsilon^2 \, ,
\end{align}
\end{widetext}
where $H(n) = \sum_{k=1}^n 1/k$. Note that $\psi_1 = \phi$ and $\psi_2 = \phi_c$. Figure~\ref{curves} plots the dependence of the $\psi_l$ on $l$ for several dimensions. Our main result~(\ref{scaleC}) follows immediately from the scaling form~(\ref{scalingForm}) by taking the derivative with respect to $K_l ( \vec{\lambda} )$ evaluated at $\vec{\lambda} = \vec{0}$.

{\em Moments vs.\ cumulants.}--Knowing the scaling behavior of the cumulants $C_\vartheta^{(l)}$ one might wonder about the corresponding moments 
\begin{align}
M_\vartheta^{(l)}  (x, x^\prime) = \left\{ \left\langle   \vartheta_{x, x^\prime} ^{2l} \right\rangle \right\}_{\mathfrak{C}}^\star \, . 
\end{align}
If the bond energy $U$ is harmonic, than one readily finds by virtue of the relation $\langle \vartheta^{2l} \rangle = (2l)!/(2^l l!) \langle \vartheta^{2} \rangle^l_c$ that
\begin{align}
\label{harmonicMoments}
M_\vartheta^{(l)}  (x, x^\prime) \sim  |x - x^\prime|^{l \phi/\nu} \, ,
\end{align}
i.e., the moments display conventional gap scaling. The situation is much more intricate if $U$ is not harmonic because then the higher moments correspond to complicated sums of products of the cumulants. We cannot prove, but it is not implausible that
\begin{align}
\label{speculation}
\bigg\{ \prod_{k=1}^\infty \left\langle   \vartheta_{x, x^\prime}^{2k} \right\rangle_c^{n_k}  \bigg\}_{\mathfrak{C}}^\star \sim  |x - x^\prime|^{\sum_{k=1}^\infty n_k \psi_k/\nu} \, ,
\end{align}
Provided that this holds, one is led back to Eq.~(\ref{harmonicMoments}) for the leading behavior of the moments in the limit $|x - x^\prime| \to \infty$ because $\psi_l$ is a strictly monotonically decreasing function of $l$ and hence $\sum_{k=1}^\infty n_k \psi_k \leq l \phi$ with $l = \sum_{k=1}^\infty n_k$.

{\em Concluding remarks.}--In summary, we have studied multifractality in broad class of systems which includes the RRN and the diluted $x$-$y$ model. The number of critical exponents $\psi_l$ required to describe the scaling behavior of the cumulants defined in Eq.~(\ref{defCphi}) corresponds to the number of terms required in a power series expansion of the bond energy $U$. In the RRN, $U$ is harmonic and hence the cumulants~(\ref{defCphi}) show no multifractality. In the diluted $x$-$y$ model infinitely many terms are required and one has true multifractality in this case. Note that only the first few $\psi_l$ differ significantly from their large $l$ limit $\psi_\infty =1$. Hence, systems where the bond energy is not harmonic but when expanded features several terms beyond harmonic order will be hard to distinguish experimentally from systems with true multifractality. One might say that these systems are effectively multifractal. The scaling behavior of the moments corresponding to the cumulants~(\ref{defCphi}) remains a challenging open problem. We hope that our work stimulates experiments or computer simulations to decide whether these moments inevitably display gap scaling or not.

Helpful discussions with H. K. Janssen and T. C. Lubensky are gratefully  acknowledged. This work was supported by the Emmy Noether-Programm of the Deutsche Forschungsgemeinschaft.



\begin{thebibliography}{}
\bibitem{fotenote_regularization}
If the dynamic variable is defined on a bound interval one naturally obtains a $D$-dimensional torus. If not, one has to restrict the replica variable to a $D$-dimensional torus for regularization purposes. In this case one sets $\vec{\theta}=\vec{k}\Delta \theta$, where $\vec{k}$ is a $D$-dimensional vector with integer components $k^{(\alpha )}$ and $-M<k^{(\alpha )}\leq M$. To recover the physical situation, on has to take the replica limit $D\rightarrow 0$, $M\rightarrow \infty$ with $M^{D}\rightarrow 1$ and $\Delta \theta =\theta _{0}/\sqrt{M}$.
\end{thebibliography}
\end{document}